\documentclass[sigconf]{acmart}
\renewcommand\footnotetextcopyrightpermission[1]{} % removes footnote with conference information in first column
\usepackage{paralist}
\usepackage{booktabs} % For formal tables
\usepackage{hyperref}
\usepackage{breakurl}

\usepackage{pifont}

\usepackage{textcomp,booktabs}
\usepackage{wrapfig}
\usepackage{floatflt}
\usepackage{caption}
\usepackage{subcaption}
\usepackage{multirow}
\usepackage{graphicx}
\usepackage[linesnumbered,ruled]{algorithm2e}

\SetCommentSty{mycommfont}

\usepackage{url}

\newcommand{\ignore}[1]{}
\newcommand{\revised}[1]{}

\usepackage{xcolor}
\usepackage{listings}
\usepackage{soul}
\usepackage{xspace}
\usepackage{balance}
\usepackage{hyperref}

 \hypersetup{
    colorlinks = true,
    linkcolor = red,
    anchorcolor = blue,
    citecolor = blue,
    filecolor = blue,
    urlcolor = blue
}

\definecolor{bittersweet}{rgb}{1.0, 0.44, 0.37}

\usepackage{moresize}
\newcommand\DejaVuttfamily{%
  \fontfamily{DejaVuSansMono-TLF}\selectfont
}
\lstdefinelanguage{mc}{
  numbers=left,
  stepnumber=1,
  breaklines=true
  rulecolor=\color{black},
  keywordstyle={[2]\color{red}},
  keywords=[2]{static, inline, volatile, return, const},
  keywordstyle={[3]\color{blue}},
  keywords=[3]{void, int, struct, char, if, else, for, uint32_t, __volatile, uint8_t, sgx_sealed_data_t, BIO, double, dA, FILE, sgx_status_t, sgx_sha_state_handle_t},
  morecomment=[l]{//},
  commentstyle=\color{bittersweet}
}

\usepackage{tikz}

\usepackage{pgfplots}

\usepackage{color}
\newcommand{\authnote}[2]{{\bf[#1: #2]}}

\usepackage{xspace}

\renewcommand{\paragraph}[1]{\vspace{0.05in}\noindent{\bf{#1}.}}

\usepackage[all]{nowidow}
 \usepackage{caption}
\usepackage{shortcuts}

\renewcommand{\paragraph}[1]{\vspace{0.05in}\noindent{\bf{#1.}}}

\setcopyright{none}
\begin{document}

\title{Towards Memory Safe Python Enclave for Security Sensitive Computation} % TODO: replace with your title
\author{Huibo Wang, Mingshen Sun, Qian Feng, Pei Wang, Tongxin Li, Yu Ding}
\affiliation{%
  \institution{Baidu Security, USA}
}
%\title{Terminating Memory Corruptions for Intel SGX Enclaves with Type Safe SGX SDKs}

% % TODO: replace this section with code generated by the tool at https://dl.acm.org/ccs.cfm
% <concept_id>10002978.10003029.10011703</concept_id>
% <concept_desc>Security and privacy~Usability in security and privacy</concept_desc>

% \ccsdesc{Security and privacy~Use https://dl.acm.org/ccs.cfm to generate actual concepts section for your paper}
% % -- end of section to replace with generated code

% \keywords{template, formatting, pickling} % TODO: replace with your keywords
\makeatletter\def\@IEEEpubidpullup{9\baselineskip}\makeatother

\begin{abstract}
Intel SGX Guard eXtensions (SGX), a hardware-supported trusted execution environment (TEE), is designed to protect security-sensitive applications. However, since enclave applications are developed with memory unsafe languages such as C/C++, traditional memory corruption is not eliminated in SGX. Rust-SGX is the first toolkit providing enclave developers with a memory-language. However, Rust is considered a Systems language and has become the right choice for concurrent applications and web browsers. Many application domains such as Big Data, Machine Learning, Robotics, Computer Vision are more commonly developed in Python programming language. Therefore, Python application developers cannot benefit from secure enclaves like Intel SGX and Rust-SGX. 
To fill this gap, we propose \sysname, which is a memory-safe SGX SDK providing enclave developers a memory-safe Python development environment. The key idea is to enable a memory-safe Python language in SGX by solving the following key challenges: (1) defining a memory-safe Python interpreter (2) replacing unsafe elements of Python interpreter with safe ones, (3) achieving comparable performance to non-enclave Python applications, and (4) not introducing any unsafe new code or libraries into SGX. We propose to build \sysname with PyPy, a Python interpreter written by RPython, which is a subset of Python, and tame unsafe parts in PyPy by formal verification, security hardening, and memory safe language. We have implemented \sysname and tested it with a series of benchmarks programs. Our evaluation results show that \sysname does not cause significant overhead.

\end{abstract}

\maketitle

\section{Introduction}
\label{sec:intro}
In current systems, there is a large attack surface, including OS, VMM (Virtual Machine Manager), and other running applications. This makes it nearly impossible to inspect code for exploitable bugs.
Intel introduced Software Guard eXtensions (SGX) in 2013 to provide applications the capability to defend their secrets through the use of enclaves. An SGX enclave is a protected environment consisting of only application code and data which is protected from malware in OS, VMM, BIOS, drivers, and other applications. SGX code and data are always encrypted to and from CPU chip to memory. The SGX enclave programming model is based on a new set of CPU-privileged instructions that provides isolation from outside agents and other enclaves. SGX also offers a software attestation scheme which allows remote agents a method to authenticate the software running inside an enclave.

SGX was initially designed to protect small segments of an application that dealt with sensitive data (e.g., credit card numbers, signing keys), and also able to provide code secrecy~\cite{bauman2018sgxelide}. The critical point is that the SGX-developed application itself still must be proven memory safe and free from vulnerabilities. Nonetheless, there would be huge benefits if legacy applications could be adapted to execute within SGX enclaves. Software developed in memory-unsafe languages such as C/C++ would be complicated to verify formally. The same traditional memory corruption vulnerabilities that exist running under OS control can also occur within the enclave. Even programming languages that are touted as memory safe (e.g., Python, Java, Go, Rust) either interface with unsafe libraries or have some features which can be exploited. There are several studies that have illustrated that memory hijacking attacks can still occur in SGX.~\cite{shacham2007geometry, lee2017hacking, biondo2018guard}. 

There has been many defense mechanisms proposed to defeat memory corruption attacks, such as Stack Canaries~\cite{cowan1998stackguard}, Data Execution Prevention (DEP)~\cite{van2004exec}, Address Space Layout Randomization (ASLR)~\cite{kil2006address}, Control Flow Integrity (CFI)~\cite{abadi2009control}, so on and so forth. However, they all are imperfect for various reasons, such as the performance cost outweighing the potential protection, incompatibility with legacy systems, relying on changes in the compiler toolchain, or requiring the access of program source-code. It is thus imperative to look for other alternatives to secure SGX enclave programs. By using memory-safe languages to develop enclave programs instead of these defense mechanisms is a better solution. 

There is a current work \sysnamerust \cite{ding2017poster, wang2019towards, wang2020building}, which solved this problem by providing Rust memory-safe languages for enclave developers. However, Rust is still relatively new and not widely adopted. Rust has recently been chosen for security-sensitive web browsers. Python ranks as one of the top three most popular languages across all existing code repositories and is gaining by the most significant percentage each year.
The application domain areas (Big Data, Machine Learning, Robotics, Computer Vision, Automation Testing) are heavily dependent on Python programming language. These application domains will never benefit from secure enclaves of SGX with only \sysnamerust alternative. Porting Python interpreter to SGX is not enough by using frameworks in the previous work ~\cite{graphene, wang2019running}. Python applications would still suffer memory corruptions since the interpreter and augmented runtime are all written in C/C++ code and thus contaminating the Python program's memory safety. 

To fill this gap, we propose \sysname, a memory-safe Python SGX SDK based on PyPy, a Python interpreter written in RPython, a subset of Python language. By taming all of the unsafe parts in PyPy using security hardening, formal verification, and memory-safe languages, \sysname provides a memory-safe Python interpreter - \mesa. Since PyPy is dependent on the C standard library, and SGX has a minimal set of libraries, the previous method added a shim layer or utilized Library OS. However, by adding this layer, the extra C/C++ unsafe code was introduced. \sysname avoids this by properly customizing \mesa. Therefore, \sysname provides a memory-safe development environment.

We have implemented \sysname and executed several benchmark programs for performance analysis. Our evaluation results show that \sysname only imposes modest overhead compared to running in native Linux. \sysname has been released as an open-source project in Nov, 2018.\footnote{The source code of \sysname has been released on GitHub at \url{https://github.com/mesalock-linux/mesapy}.}.

In summary, our contributions are:
\begin{compactitem}
\item We present \mesa, a memory-safe Python interpreter, based on PyPy by using formal verification, type system enhancement, and memory safe programming language to tame the unsafe parts in PyPy. 

\item We propose \sysname based on \mesa, a practical approach to providing a memory-safe Python development environment for developers.

\item We have implemented \mesa, \sysname, and evaluated them with several benchmark programs. Our evaluation results show that \sysname provides a prototype of the Python enclave programming model and meanwhile does not impose any significant performance overhead.
\end{compactitem}

\section{Background}
\label{sec:back}
\subsection{Memory Safety}
\label{sub:safety}
Computer programs consist of two main elements: execution control and memory data access. 
Memory data access can have either spatial and temporal problems. Spatial errors are when you can read or write to memory areas that should not be valid, which includes buffer overflows, double frees, and dangling pointers. Temporal errors are the result of memory access that should not be valid because of timing issues such as dereferencing pointer after free, uninitialized memory access. The whole idea of memory safety is to provide a secure execution environment void the types of spatial and temporal issues just presented.

\subsection{Intel SGX}
The main reason for Intel to introduce SGX is to provide applications the ability to execute code in a secure enclave and protect secrets with their own execution environment~\cite{intel:sgx-intro,sgxprogreference}. As such, SGX provides software developers direct control over their application's security without relying on any underlying system software such as the OS or hypervisor. This significantly reduces the trusted computing base (TCB) to the smallest possible footprint (only the code executed inside the enclave), and prevents various software attacks even when the OS, BIOS, or hypervisor are compromised. The main reason for Intel to introduce SGX is to provide applications the ability to execute code in a secure enclave and protect secrets with their own execution environment~\cite{intel:sgx-intro,sgxprogreference}. As such, SGX provides software developers direct control over their application's security without relying on any underlying system software such as the OS or hypervisor. This significantly reduces the trusted computing base (TCB) to the smallest possible footprint (only the code executed inside the enclave) and prevents various software attacks even when the OS, BIOS, or hypervisor are compromised. 

\subsection{PyPy}
\label{sub:languageP}
PyPy is an alternative implementation of the Python programming language, which often executes faster than the standard implementation of Python, CPython. PyPy uses just-in-time (JIT) compilation to translate Python code into machine-native assembly language while CPython is strictly an interpreter. PyPy uses optimization techniques found in other just-in-time compilers for dynamic languages. It analyzes running Python programs to determine the type of information of objects as they are created and used in programs, then uses that type of information as a guide to speed things up.

"Pure" Python applications which do not interface with C libraries or extensions execute most efficiently with PyPy. This is due to the overhead incurred by how PyPy must emulate CPython’s native binary interfaces. Longer-running programs benefit most from PyPy optimizations. The longer the program runs, the more run-time type information PyPy can gather, and the more optimizations it can make. Python used for system command scripting purposes will not benefit from the efficiency standpoint.

PyPy is interesting architecturally because it is the product of a technique called meta-tracing, which transforms an interpreter into a tracing JIT compiler. Since interpreters are usually easier to write than compilers but run slower, this technique can make it easier to produce efficient implementations of programming languages. PyPy's meta-tracing toolchain is called RPython. PyPy compiles Python code, but it is not a compiler for Python code. Because of the way PyPy performs its optimizations and the inherent dynamism of Python, there is no way to emit the JIT-generated byte code as a standalone binary and re-use it.

RPython is a Python-like language; specifically it is a restricted subset of the Python language. The restriction provides the power of type inference of the RPython program so that it could be translated into another language. The mechanism to achieve this byte code translation is through an associated toolchain provided by PyPy project along with an interpreter written in the C programming language. This code is then compiled into machine code, and the byte code runs on the compiled interpreter.

\section{Overview}
\label{sec:overview}
\subsection{Objectives, Threat Model and Scope}
The critical objective of \sysname is to provide a secure SGX enclave environment for Python software development free of memory-corruption vulnerabilities. \sysname should provide to the most extent possible the same capabilities and rich library support that is found in the native Linux environment. Applications should not have to undergo a significant rewrite to execute in \sysname. Moreover, finally, the design and implementation of \sysname should not impose significant performance overhead.

\sysname, like Intel SGX, only protects code and data running from within the enclave itself. The CPU processor protects code running in the enclave from being "spied on" by other code, including OS processes, BIOS, hypervisor, or other applications. The enclave contents are unable to be read by any code outside the enclave, other than in encrypted format. This SGX model works securely and correctly as long as there are no memory safety issues (buffer overflow, use after free, invalid pointer). SGX programs are at risk of the same control flow hijacking as traditional applications when these memory safety rules are violated. Making matters more complicated, an SGX application is normally implemented with a trusted component running inside the enclave and an untrusted component running outside as an application software process. Since data and control must be passed back and forth between bridge functions (ECALL and OCALL), this opens another window of opportunities for attack. \sysname aims to defeat memory corruptions that exploit the insecure memory operations inside enclave programs. Side-channel attacks are orthogonal to the memory safety problem, so they are not in scope. \sysname will only focus on enabling application layer memory safety in SGX enclaves.   

\subsection{Challenge and Threat}
\begin{figure}
\begin{center}
    \includegraphics[width=\linewidth]{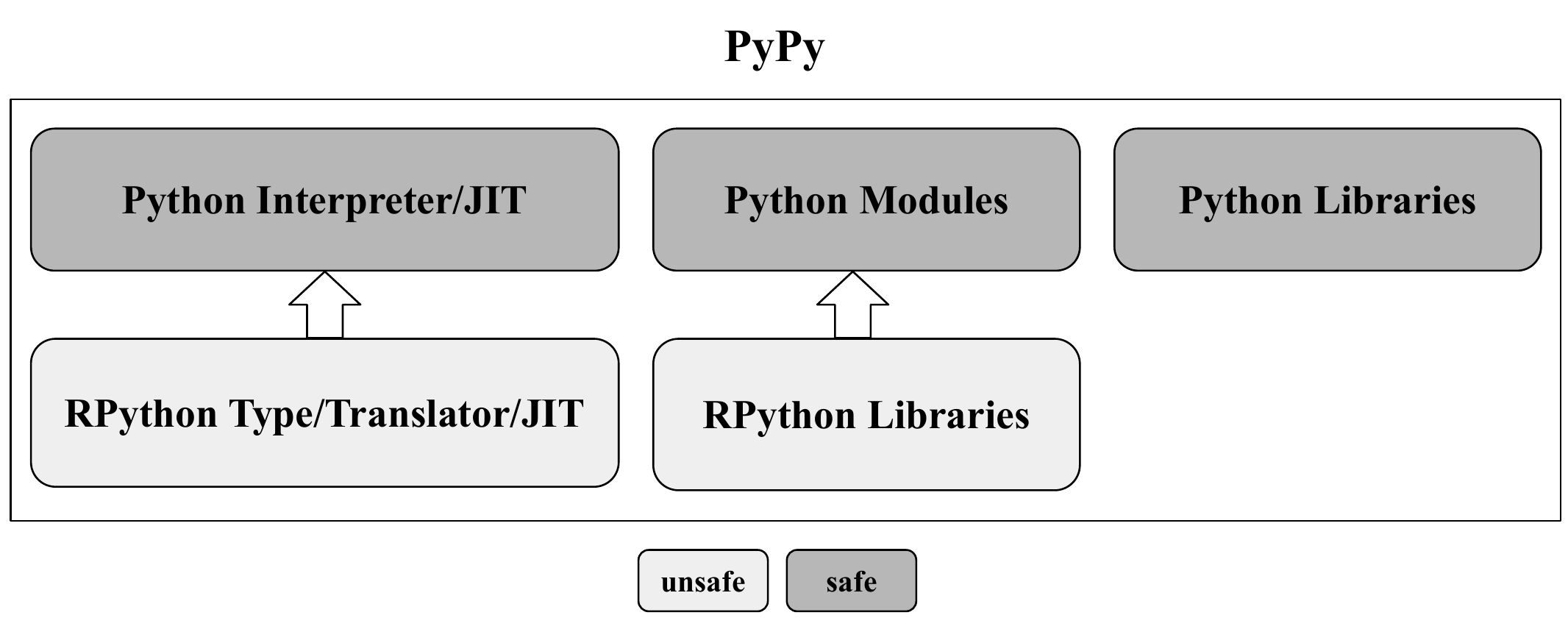}
\end{center}
  \caption{PyPy Structure.}
   \label{fig:pypyarch}
\end{figure}

In the previous work related to SGX memory safety, \sysnamerust focused on solving the unsafe interface between Rust and C languages. \sysname focuses on solving the memory safety of scripting language itself, in particular, its interpreter. Although Python is a type-safe scripting language, Python has to depend on an interpreter to execute. Python applications would still suffer memory corruption since the interpreter is not safe. The most popular Python interpreter is CPython, which is written in C, which is an unsafe language. Therefore, the first challenge is how to make Python's interpreter memory safe. One method will be using formal verification tool to find and correct any potential memory corruptions in CPython; the other one will be rewriting CPython with memory-safe language. For the first one, CPython has 300k lines of C code, which will be a near-impossible task to accomplish. The second one comes to our solution. There is a Python interpreter, PyPy, which is written in RPython, which is a subset of Python. However, in PyPy, there are three unsafe parts, all related to RPython: type system, Translator/JIT backend, and external libraries. PyPy structural components are illustrated in ~\autoref{fig:pypyarch}. Therefore, the challenge finally comes to how to secure these unsafe parts in PyPy interpreter. \sysname addresses this issue by using security hardening, formal verification, and memory safe programming language to eliminate the unsafe parts in the interpreter. The new memory-safe PyPy is called \mesa, which can also be used in Linux environments. 

The second challenge is the same one faced by \sysnamerust. \sysname is built on top of standard C library, not Intel SGX SDK. Since Python interpreter depends on the standard C library, it is impractical to have every component inside the SGX enclave to be memory safe. Therefore, \sysname will need to be re-designed to build on top of Intel SGX SDK. 

Since PyPy depends on C standard library, and SGX can only utilize a subset of this library interface (I/O is forbidden, no system calls, etc.), PyPy can not be executed in SGX.%\sysname will have limited programming functionality. 
The previous method was to add a shim layer or use Library OS. However, by adding this layer, the extra C/C++ unsafe code was introduced. Therefore, properly porting Python interpreter in Intel SGX without introducing any extra memory unsafe parts comes to another key challenge. \sysname addresses this issue by analyzing the mismatch between interpreter and SGX limited programming environment. This includes avoiding any extra C/C++ layer, following the SGX requirement, and modifying the interpreter, so it works correctly under SGX limited functionality.

\section{Design and Implementation}
\label{sec:design}
\begin{figure}

\begin{center}
    \includegraphics[width=\linewidth]{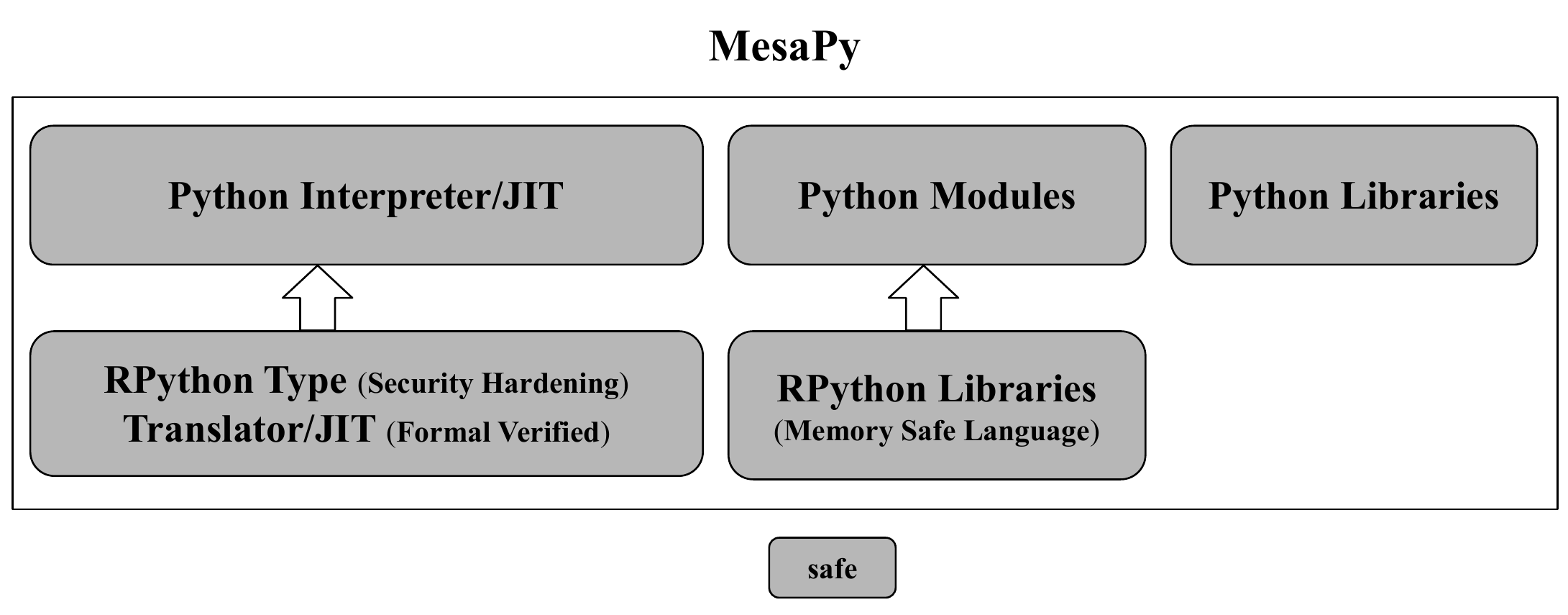}
\end{center}
  \caption{Overview of \mesa.}

  \label{fig:mesapy}
\end{figure}

PyPy, a Python interpreter written primarily in RPython \S\ref{sub:languageP}, has built-in memory safety and speed efficiency as a result of its JIT compiler feature. However, there are still three unsafe parts which are: (1) RPython Type System, (2) RPython's Translator/JIT backend, and (3) RPython's external libraries. In the following sections, we introduce solutions to ensure safety in these three areas shown in \autoref{fig:mesapy} and securely porting of \mesa in SGX. 

\subsection{Unsafe RPython Type System - Security Hardening}
For complete memory safety, both spatial (out of bound) and temporal (use-after-free) errors must be prevented without false negatives. Type-safe languages usually enforce both spatial and temporal safety by checking object bounds at array accesses and using automatic garbage collection. 

RPython utilizes garbage collection to prevent temporal safety. However, by auditing RPython implementation in PyPy, we found that RPython has a potential risk since it does not check the boundaries of list and arrays, which will break spatial safety. We need to generate exceptions for unsafe functions operating outside the legal boundaries of the list and arrays. By performing security hardening on RPython's type system, we improve RPython's spatial safety.

\subsection{Unsafe RPython's Translator/JIT backend}
There are approximately 1000 lines of runtime C code in RPython's Translator/JIT backend, which contain potential memory bugs. %Guaranteeing code execution safety is another part of PyPy memory safety. 
These few lines of C code are difficult to eliminate since they are often referenced.
Although testing can be very effective at finding bugs, it cannot uncover all issues because exhaustively enumerating all program inputs is not possible. Therefore, formal verification is a complementary approach that can be effectively used to prove the absence of bugs in addition to finding them. Since this C code is isolated, its logic is not complicated and it is a critical piece, we can deploy formal verification methods to guarantee four memory issues cannot occur: (1) buffer overflow, (2) buffer over-read, (3) null pointer dereference, and (4) memory leak.

Two approaches are used for formal verification. The first one is the Abstract Interpretation (AI) based verification. Abstract Interpretation is a framework of program analysis based on approximating a possibly infinite set of states into a single abstract state. This leads to a computable set of states representing safe approximations of the behavior of the program. The second approach is using Software Model Checking (SMC). SMC is a method that verifies the correct functioning of a system design model typically represented by finite state machines (FSM). SMC uses efficient algorithms to verify that execution behavior adheres to formal specification and creating counterexamples for any violations. SMC is more robust with more complex control-driven programs with a large number of finite states. AI is more robust with data-dependent programs since it performs more exhaustive verification of abstract value space.

%tries to provide an abstract mathematical model of the program, and prove the absence of bugs in it. 

In practice, we deployed tools that cover both approaches in order for cross-validation. We utilize two state-of-the-art verification tools -- SeaHORN~\cite{seahorn} and SMACK~\cite{cav2014-re}
%, and TIS (TrustInSoft)~\cite{tis} 
to reach a high degree of certainty that memory safety issues are eliminated. Seahorn and SMACK both encompass aspects of SMC and AI. Seahorn employs several SMC-based model checking engines (PDR/IC3, SPACER, and GPDR) as well as AI-based analyzer IKOS for checking program invariants. SMACK is well known for SMC bounded verification, but it can also serve AI verification in programs adequately annotated with loop invariants and procedure pre- and post-conditions. Both tools use LLVM intermediate representation (IR) as the front end. C/C++ code is translated to LLVM IR bitcode first and then processed by the middle-end process. Seahorn takes LLVM IR bitcode and emits Contrained Horn Clauses (CHC) for the back-end process. SMACK converts LLVM IR bitcode to Boogie IVL, which is an expressive mathematical language based on memory map regions instead of dynamic memory. Finally, both tools have back-end verification engines to fulfill program validation.

The four aspects (buffer overflow, buffer over-read, null pointer dereference, and memory leak) are the verification target scopes. We applied both tools on the RPython backend C programs. To evaluate the verification result, we manually checked each of the alarms reported by both tools. By looking into the intermediate results, mapping the semantics to the source code manually, we can prove the results of back-end verification engines.

By formal verification, we found several cases could be potential vulnerabilities. Here, we list two cases to show the verification results.

\begin{figure}
\centering
\footnotesize
\begin{lstlisting}[language=mc,
basicstyle=\footnotesize\ttfamily]
static Bigint *
Balloc(int k)
{
    int x;
    Bigint *rv;
    unsigned int len;
    if (0<=k <= Kmax && (rv = freelist[k]))
        freelist[k] = rv->next;
    else {
        x = 1 << k;  // ---> potential integer overflow
       ....... code omit
         }
}
\end{lstlisting}
\caption{vulnerability in dtoa.c}
\label{lst:dtoa}
\end{figure}

\begin{figure}
\centering
\footnotesize
\begin{lstlisting}[language=mc,
basicstyle=\footnotesize\ttfamily]
static Bigint *
Balloc(int k)
{
    int x;
    Bigint *rv;
    unsigned int len;
    if (0<=k <= Kmax && (rv = freelist[k]))
        freelist[k] = rv->next;
    else {
        if(k>32 || k<0) return NULL; //---> prevent integer overflow
        x = 1 << k;  
       ....... code omit
       }
}
\end{lstlisting}
\caption{Protection in dtoa.c}
\label{lst:dtoafix}
\end{figure}

\begin{figure}
\centering
\footnotesize
\begin{lstlisting}[language=mc,
basicstyle=\footnotesize\ttfamily]
RPY_EXTERN
long *pypy_jit_codemap_del(unsigned long addr,
                            unsigned int size)
{
    unsigned long search_key = addr + size - 1;
    long *result;
    skipnode_t *node;
    node = skiplist_search(&jit_codemap_head, search_key); 
    pypy_codemap_invalid_set(1);
    skiplist_remove(&jit_codemap_head, node->key);  // --> invalid memory access
    ....... code omit
}
\end{lstlisting}
\caption{vulnerability in codemap.c}
\label{lst:codemap}
\end{figure}

In \autoref{lst:dtoa}, there could be a potential Integer Overflow at line 10 since K could be any value, including these greater than 32. In such a case, x will be overflowed.

In \autoref{lst:codemap}, A potential invalid memory access could happen at line 10. 
Since \texttt{skiplist-search} could return NULL, \texttt{node} could be NULL, thus \texttt{node->key} will cause a crash.

We fixed all tool detected potential memory corruptions manually one by one and then iteratively revalidated all code updates against the verification tools until no potential risks existed. The correction of potential corruptions in ~\autoref{lst:dtoa} and ~\autoref{lst:codemap} are shown in ~\autoref{lst:dtoafix} and ~\autoref{lst:codemapfix}.

\begin{figure}
\centering
\footnotesize
\begin{lstlisting}[language=mc,
basicstyle=\footnotesize\ttfamily]
RPY_EXTERN
long *pypy_jit_codemap_del(unsigned long addr,
                            unsigned int size)
{
    unsigned long search_key = addr + size - 1;
    long *result;
    skipnode_t *node;
    node = skiplist_search(&jit_codemap_head, search_key);
    if (!node || !node->key || node->key < addr)
    return NULL;//--> check if node is null ptr
    pypy_codemap_invalid_set(1);
    skiplist_remove(&jit_codemap_head, node->key);  
    ....... code omit
}
\end{lstlisting}
\caption{protection in codemap.c}
\label{lst:codemapfix}
\end{figure}

\subsection{Unsafe RPython external libraries}
\label{sub:lib}
In RPython, there are external libraries written in C/C++, such as zlib, openssl, expat, etc. These external libraries written in C/C++ can have potential memory issues. Since it is hard to audit the libraries depending on the size of libraries, how can memory safety be guaranteed? There are two solutions for these unsafe external libraries; one is to use memory safe programming languages to rewrite them, the other one is to use formal verification to verify the correctness. However, external libraries are often not scalable and have a large amount of complicated code and dependencies. It is impractical to use formal verification to achieve memory safety. For unsafe external libraries, using memory safe programming language to rewrite them is a better approach. In practice, We replaced zlib with miniz-oxide, which a minizlib was written by Rust.

\subsection{Porting PyPy into SGX}
SGX reduces the size of the required Trusted Computing Base (TCB) significantly. However, this restrictive system model also limits SGX programs' capability of acquiring computation resources from the outside untrusted environment, including hardware and software interrupts. These features are disabled in SGX enclaves, and their associated services from the OS, I/O subsystem, and memory mapping are unavailable. 

Since SGX has limited library support, the key challenge is having PyPy run correctly inside SGX with the minimal TCB while not introducing unsafe code in SGX. The SGX security guide must be followed during the development of new libraries.

PyPy interpreter has RFFI parts which contain system calls and I/O related operations. This introduces environment variables, disk files, dynamic load, etc. which are not supported by SGX. The challenge becomes how to fix these I/O and system call-related requirements.

Previous works have focused on building a bridge for the limitations. However, it increases the TCB and also adds an extra layer inside SGX, which is written in C/C++ language introducing more potential memory corruption.

Instead, the PyPy interpreter has been modified directly for \sysname without adding an unsafe C/C++ code. Also, we performed analysis for all I/O and system call functions used by PyPy, so that they could either be eliminated or replaced with equivalent safe implementation. For instance, there is a requirement to read files from disk in PyPy for booting. As an alternative, we can have these files prepared inside SGX for reading. Since PyPy interpreter no longer requires a dynamic loading feature inside SGX, we can disable this function and replace it with static loading. By searching through RPython FFI, we remove all of the I/O and system calls that are not allowed. 

In order to provide more details on this practice, let us examine \texttt{pypy-setup-home} function which implements PyPy interpreter setup and performs \texttt{lib-python} and \texttt{lib-pypy} initialization. During this process, these libraries need to read files from disk, which is outside of SGX. In order to remove potential risk from reading files from an unsafe source, we can hardcode these two libraries inside SGX. Moreover, since PyPy is also supported by libc, which contains unsafe code, all libc functionality will be replaced with sgx tlibc. 

As another example, garbage collection initialization needs to read system information about the total memory size. This part is redesigned so that there is no need to read from the file system by defining the size accurately. Further analysis is performed for unnecessary code, which may contain unsafe actions reducing the threat surface area. PyPy interpreter then becomes self-contained in SGX and isolated from potential risks.

% \section{Implementation}
% \label{sec:implementation}
% \input{paper/5_implementation.tex}

\section{Evaluation}
\label{sec:eval}
In this section, we provide a performance evaluation of \sysname. Two areas are identified as the criteria for performance evaluation: (1) \mesa versus PyPy (both running in Linux), and (2) porting overhead (\mesa in Linux versus \mesa in \sysname). Note, we do not compare a Python program with a C/C++ program since the comparison is inappropriate. That would necessarily be the comparison compiled versus interpreted code. Since we have not developed any SGX-specific APIs written in Python yet, we will also not have micro-benchmarks to evaluate that part yet.
The source code of all of our benchmarks is released at Github at \url{https://github.com/mesalock-linux/mesapy-benchmarks}.

Our performance experiments were executed on a 16.04.1-Ubuntu system comprised of a 6-core Intel Core i7-8086K CPU running at 4.00GHz, with 32 GB memory, and 1Gbit/s Ethernet Connection I219-LM. We installed the latest Intel SDK, SGX driver(version 2.5).

%\subsection{\sysname Performance Evaluation}
\subsection{\mesa benchmarks}
Our benchmarks in this part aim to evaluate the overhead of \mesa compared to native PyPy. 
Particularly, we set up two sets of evaluation environments: (1) PyPy with Linux, 2) \mesa with Linux. We have 19 PyPy~\cite{pypy} and game-related programming~\cite{gamebenchmark} benchmarks that are selected and modified for execution. The evaluation result for these benchmarks is presented in \autoref{fig:pypy}. We can see that all 16 programs running in \mesa have the overhead ranging from 1\% to 12\% greater than native PyPy. The reason that native PyPy has better performance than \mesa is due to the list and array index check in \mesa. There are three benchmarks\texttt{}, \texttt{} and \texttt{} running on \mesa are slightly faster than native PyPy. The reason is mainly from noise. 

\subsection{Porting overhead}
Our benchmarks in this part aim to evaluate the overhead of porting \mesa in \sysname compared to native \mesa in Linux. We set up two sets of evaluation environments: 1) running with \mesa in Linux, 2) running under \sysname. We measured the execution time by utilizing the operating system clock running outside of the enclave. In particular, we first start the clock outside an enclave, execute ECALL, which will call a Python function inside the enclave, and then stop the clock to calculate the total execution time. 

We have eight benchmarks that are supported by \mesa under both Linux and SGX execution environments. Note that due to the limitation of SGX memory, we run performance testing with tree depth as 18 in \texttt{binary-tree}, and have 10 as input for testing \texttt{fasta}, and execute \texttt{pidigits} with input as 1.
The results are shown in \autoref{fig:mesa}.  The benchmarks reporting performance overhead is between 36\% to 150\%  higher for \mesa under \sysname compared to \mesa under Linux.  The overhead is introduced by SGX and \mesa porting. 
\pgfplotstableread{
benchmark	native	native-min	native-max	sgx-shield	sgx-shield-min	sgx-shield-max	
binary-trees	100 0	0	105	0	0	
chaos	100	0	0	101	0	0	
crypto\_pyaes	100	0	0	101	0	0	
deltablue	100	0	0	107	0	0	
fannkuch	100 0	0	112	0	0	
fasta	100	0	0	112	0	0	
fib	100	0	0	100	0	0	
float	100	0	0	99	0	0	
gcbench	100 0	0	107	0	0	
hexiom2	100	0	0	105	0	0	
json\_bench	100	0	0	106	0	0	
mandelbrot	100	0	0	102	0	0	
meteor-contest	100 0	0	100	0	0	
nbody	100	0	0	99	0	0	
nqueens	100	0	0	107	0	0	
pidigits	100	0	0	108	0	0	
schulze	100	0	0	100	0	0	
spectral-norm	100	0	0	99	0	0
telco	100	0	0	103	0	0
}{\data}

\tikzset{align at top/.style={baseline=(current bounding box.north)}}

%%%%%%%%%%%%%%%%%%%%%%%%%%%%%%%%%%%%%%%%%%%%%%%%%%%%%%%%%%%%%%%%%%%

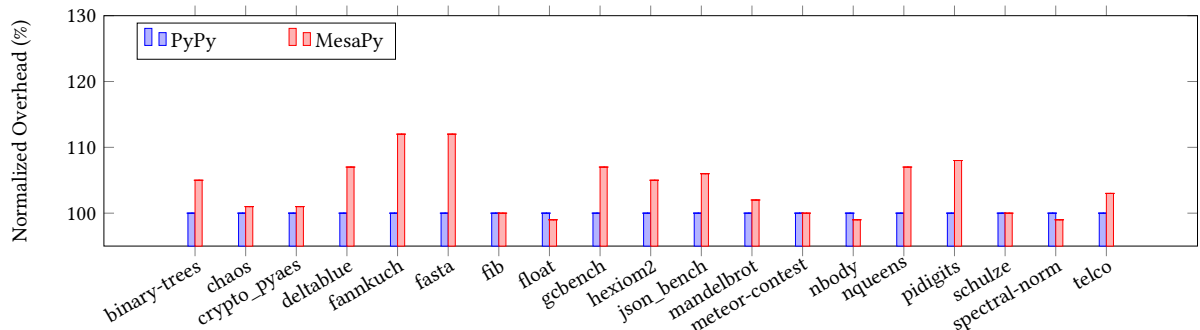
\begin{figure*}[t]
\centering
%\captionsetup{margin=0.2cm}
\scalebox{1.0}{
  
\begin{minipage}{1.0\linewidth}
  \centering
  \resizebox{0.9\linewidth}{!}{
   \begin{tikzpicture}
\begin{axis} [
    ymin=95,
    ymax = 130,
    ybar,
    height = 5.0 cm,
    ybar=0pt,
    try min ticks=5,
    width = 1.0\textwidth,
    legend style={
                	{font=\normalsize},
                    at={(0.0,-0.15)},
                    anchor=north west,
                    legend columns=-1,
                    /tikz/every even column/.append style={column sep=1.0cm}
                 },
    %width = 1\textwidth,%=8pt, % configures ‘bar shift’         
    symbolic x coords={binary-trees,
     				chaos,
                	crypto\_pyaes,
                    deltablue,
                    fannkuch,
                    fasta,
                    fib,
                    float,
                    gcbench,
                    hexiom2,
                    json\_bench,
                    mandelbrot,
                             meteor-contest,
                    nbody,
                    nqueens,
                    pidigits,
                    schulze,
                    spectral-norm,
                    telco},
    xticklabel style={rotate=30, anchor=north east, inner sep=0mm},
    legend pos= north west, 
    bar width=3pt,
    ybar=0pt,
    ylabel={Normalized Overhead (\%)},
    xtick=data
]
    
  \addplot
  plot [error bars/.cd, y dir=both, y explicit]
  table [x=benchmark, y=native,y error plus=native-max, y error minus=native-min] {\data};
    
  \addplot 
  plot [error bars/.cd, y dir=both, y explicit]
  table [x=benchmark, y=sgx-shield,y error plus=sgx-shield-max, y error minus=sgx-shield-min] {\data};
  
   \legend{PyPy, \mesa}
\end{axis} 
\end{tikzpicture}
    }
    \vspace{-0.16in}
    \caption{Percentage overhead of running Python benchmarks in \mesa normalized against PyPy native execution.}
    \label{fig:pypy}
  \end{minipage}%

  }
\end{figure*}
\pgfplotstableread{
benchmark	native	native-min	native-max	sgx-shield	sgx-shield-min	sgx-shield-max	
binary-trees	100 0	0	200	0	0	
nbody	100	0	0	199	0	0
fasta	100	0	0	230	0	0
fannkuch	100 0	0	225	0	0
pidigits	100	0	0	136	0	0	
deltablue	100	0	0	250	0	0	
fib	100	0	0	191	0	0	
nqueens	100	0	0	177	0	0
}{\data}

\tikzset{align at top/.style={baseline=(current bounding box.north)}}

%%%%%%%%%%%%%%%%%%%%%%%%%%%%%%%%%%%%%%%%%%%%%%%%%%%%%%%%%%%%%%%%%%%

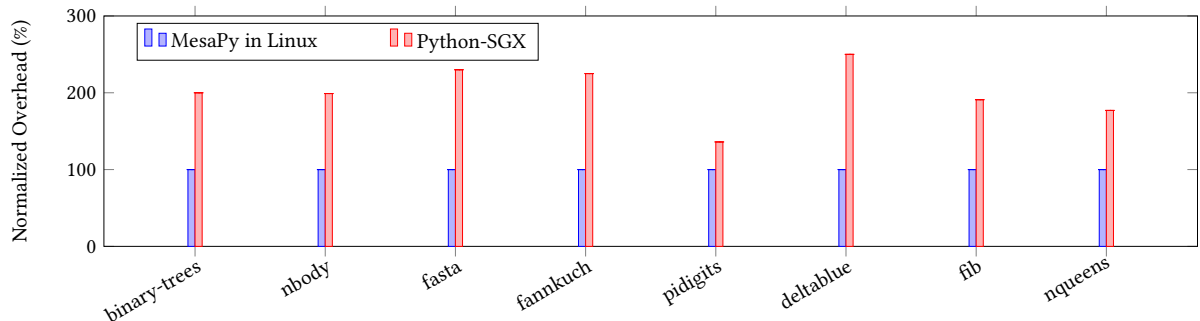
\begin{figure*}[t]
\centering
%\captionsetup{margin=0.2cm}
\scalebox{1.0}{
  
\begin{minipage}{1.0\linewidth}
  \centering
  \resizebox{0.9\linewidth}{!}{
   \begin{tikzpicture}
\begin{axis} [
    ymin=0,
    ymax = 300,
    ybar,
    height = 5.0 cm,
    ybar=0pt,
    try min ticks=5,
    width = 1.0\textwidth,
    legend style={
                	{font=\normalsize},
                    at={(0.0,-0.15)},
                    anchor=north west,
                    legend columns=-1,
                    /tikz/every even column/.append style={column sep=1.0cm}
                 },
    %width = 1\textwidth,%=8pt, % configures ‘bar shift’         
    symbolic x coords={binary-trees,
     				nbody,
                    fasta,
                    fannkuch,
                    pidigits,
                    deltablue,
                    fib,
                    nqueens
                    },
    xticklabel style={rotate=30, anchor=north east, inner sep=0mm},
    legend pos= north west, 
    bar width=3pt,
    ybar=0pt,
    ylabel={Normalized Overhead (\%)},
    xtick=data
]
    
  \addplot
  plot [error bars/.cd, y dir=both, y explicit]
  table [x=benchmark, y=native,y error plus=native-max, y error minus=native-min] {\data};
    
  \addplot 
  plot [error bars/.cd, y dir=both, y explicit]
  table [x=benchmark, y=sgx-shield,y error plus=sgx-shield-max, y error minus=sgx-shield-min] {\data};
  
   \legend{\mesa in Linux, \sysname}
\end{axis} 
\end{tikzpicture}
    }
    \vspace{-0.16in}
    \caption{Percentage overhead of running Python benchmarks in \sysname normalized against \mesa in Linux.}
    \label{fig:mesa}
  \end{minipage}%

  }
\end{figure*}

\ignore{
\subsection{Formal Verification}
All formal methods are faced with challenge of approximating runtime behavior of programs with infinite or extremely large input variable space to be solved in reasonable timeframe with finite computer system. The basic idea of formal verification is that a possible infinite set of states can be approximated by one abstract state. Abstract interpretation leads to a computable state at any time during execution of the program which provides model checking capability. Coupled with AI analysis engine, runtime error checks, and weakest precondition verification, the majority of memory safety issues can be identified.

Three formal verification tools (Smack, Seahorn, TIS) are deployed for cross verification purpose. Smack can work on C programs and LLVM bitcode. To facilitate the RPython verification, we created a Makefile to automatically generate LLVM bc output for all C source files. In Seahorn, we provided the scripts to automatically evaluate RPython Translator/JIT backend programs. The shell scripts we provided can be used to automatically generate bc and provide the Seahorn verification. TIS performs the static analysis verification on the source code. TIS not only checks the target C program source code, but also all included header files. We also need to setup magic quadrants in order for TIS to accurately perform static verification. We show results for all three methods of verification in \autoref{table:TIS}.

In the table, every verification tool will output their verification result as \textttsat{sat}(\xmark) or \texttt{unsat} (\cmark). \texttt{sat} means that a violation has been detected by memory safety checking rule. In other words, it means it is unsafe for the target program. \texttt{unsat} means that there is no violation for the memory safety check. It means the target program is memory safe.
We can see the results from these three verification tools. Some memory safety bugs are not detected from each tool. However, by using cross verification, we can cover more code. We fix all of unsafe code manually one by one.

\begin{table*}[t]
\caption{Formal Verification Results}
\begin{tabular}{ccccccc}
\toprule
\textbf{Filename} & \textbf{LOC} & \textbf{TIS Results} & \textbf{Smack Results} & \textbf{Seahorn Results}\\
\midrule
dtoa.c          & 3006 & \xmark & \xmark   & \cmark     \\ 
asm\_gcc\_x86.c & 33 & \xmark   & \cmark   & \cmark  \\ 
entrypoint.c    & 130 & \cmark  & \cmark   & \cmark   \\ 
codemap.c       & 165 & \xmark   & \xmark  & \xmark    \\ 
skiplist.c      & 109 & \xmark   & \cmark   & \cmark   \\ 
debug\_print.c  & 274 & \cmark   & \cmark  & \xmark   \\ 
exception.c     & 45 & \cmark    & \cmark  & \cmark  \\ 
instrument.c    & 76 & \cmark    & \cmark  & \cmark  \\ 
int.c           & 48 & \cmark    & \cmark  & \cmark \\ 
ll\_strtod.c    & 145 & \xmark  & \xmark    & \xmark   \\ 
mem.c           & 178 & \xmark  & \xmark    & \xmark   \\ 
profiling.c     & 73 & \cmark   & \xmark    & \xmark \\ 
rtyper.c        & 38 & \xmark    & \xmark   & \xmark  \\ 
signals.c       & 206 & \cmark   & \xmark   & \xmark  \\ 
stack.c         & 66 & \xmark    & \xmark   & \xmark  \\ 
support.c       & 27 & \cmark    & \xmark   & \xmark  \\ 
thread.c        & 22 & \cmark    & \xmark   & \xmark  \\ 
threadlocal.c   & 273 & \cmark   & \xmark   & \xmark  \\ 
\bottomrule
\end{tabular}
\label{table:TIS}
\end{table*}
}

\section{Discussion}
\label{sec:discussion}
\subsection{Limitation of current work.}
In the current version, \sysname only supports basic computation and some builtin modules. Support for multi-threading and full standard library is still under development.

\subsection{The future work.} 
There is work remaining to perfect \sysname, including porting more Python libraries into \sysname and replacing remaining external C libraries. Furthermore, we still need to verify more components using current state-of-the-art verification tools.
To have \sysname adopted by a larger developer community, we will be implementing additional useful libraries and SGX specific features for real-world use cases.

The audit of translation from RPython to C will be included in the next scope. 

\section{Related Work}
\label{sec:related}
There exist many defense-related works regarding memory safety.
G-Free is a work defeating return-oriented programming through gadget-less binaries~\cite{onarlioglu2010g}.
Hypersafe is a lightweight approach to provide lifetime hypervisor control-flow integrity~\cite{wang2010hypersafe}.
Pointguard TM is protecting pointers from buffer overflow vulnerabilities~\cite{cowan2003pointguard}.
CCured is doing type-safe retrofitting of legacy code~\cite{necula2002ccured}.
~\cite{ruwase2004practical} presents a Practical Dynamic Buffer Overflow Detector.
Baggy Bounds Checking~\cite{akritidis2009baggy} is an Efficient and Backwards-Compatible Defense against Out-of-Bounds Errors.
~\cite{yong2003protecting} is protecting C programs from attacks via invalid pointer dereferences.
~\cite{jones1997backwards} presents backward-compatible bounds checking for arrays and pointers in C programs.
Cling~\cite{akritidis2010cling} is using a Memory Allocator to Mitigate Dangling Pointers.
CETS~\cite{nagarakatte2010cets} presents compiler enforced temporal safety for C.
~\cite{akritidis2008preventing} is preventing memory error exploits with WIT.
Body Armor for Binaries~\cite{slowinska2012body} is preventing Buffer Overflows Without Recompilation.
Stackguard~\cite{cowan1998stackguard} presents automatic adaptive detection and prevention of buffer-overflow attacks.

Since the root cause of all vulnerabilities is memory corruption,
every exploit starts by triggering a memory error. Enforcing memory safety eliminates all memory corruption exploits. While achieving memory safety, both spatial and temporal errors must be prevented without false negatives. Type-safe languages enforce both spatial and temporal safety by checking object bounds at array accesses and using automatic garbage collection. We can transform existing unsafe code to enforce similar policies by embedding low-level reference monitors. This instrumentation can appear within the source code itself, intermediate representation, or at the binary level. We should solve this problem for both spatial and temporal safety~\cite{szekeres2013sok}. 

\paragraph{Spatial safety with pointer bounds}
Enforcing spatial safety is to keep track of pointer bounds. CCured~\cite{necula2002ccured} provides an extension of the C type system with explicit types for pointers into arrays with dynamic typing.  Cyclone~\cite{jim2002cyclone} use "fat pointers". However, these systems need source code annotations so that it is not practical for large codebases. Also, any pointer representation changes the memory layout and breaks binary compatibility. SoftBound~\cite{nagarakatte2009softbound} addresses the compatibility problem of pointer representation by separating metadata from the pointer.

% \section{Availability}
% \label{sec:availability}
% \input{paper/09_availability.tex}

\section{Conclusion}
\label{sec:conclusion}
In summary, we have presented \sysname, a framework based on a safely ported PyPy interpreter to SGX providing enclave developers a memory-safe Python development environment. We show that with \sysname, enclave developers can perform sensitive security computations without significant overhead. 

In order to be more practical, we plan to bring in additional Python libraries and SGX-specific features for real-world use cases. We also plan to review the translation audit from RPython to C for increased memory safety evaluation.

\bibliographystyle{IEEEtranS}
\bibliography{references.bib}

\end{document}